\documentclass[12pt]{article}   
\usepackage{epsfig}
\usepackage[utf8]{inputenc} 
\usepackage{url}
\usepackage{here}
\usepackage{amssymb}
\usepackage{graphicx}
\usepackage{comment}
\usepackage{multirow}
\usepackage{verbatim}
\usepackage{fancyvrb}

\begin{document}


\title{On a curious bias arising when\\ the $\sqrt{\chi^2/\nu}$
  scaling prescription is first applied\\ to a sub-sample
  of the individual results}
\author{Giulio D'Agostini \\
Universit\`a ``La Sapienza'' and INFN, Roma, Italia \\
{\small (giulio.dagostini@roma1.infn.it,
 \url{http://www.roma1.infn.it/~dagos})}
}
\date{}
\maketitle

\thispagestyle{empty}

\begin{abstract}
  As it is well known, the standard deviation of a weighted average
  depends only on the individual standard deviations, but not
  on the  dispersion of the values around the
  mean. 
  This property  leads sometimes to the
  embarrassing situation in which the combined result
  `looks' somehow at odds with the individual ones.
  A practical way to cure the problem is to enlarge
  the resulting standard deviation by the 
  $\sqrt{\chi^2/\nu}$ scaling, a {\em prescription} employed 
  with arbitrary criteria on when to apply it and
  which individual results to use in the combination.
  But the `apparent' discrepancy between the combined result
  and the individual ones often remains. Moreover this rule does not
  affect the resulting `best value', even if the pattern of the
  individual results is highly skewed. 
  In addition to these reasons of dissatisfaction, shared by many practitioners,
  the method causes another issue,
  recently noted on the published measurements
  of the charged kaon mass. 
  It happens in fact that, if the prescription is applied twice,
  i.e. first to a sub-sample of the individual results and subsequently
  to the entire sample, then {\em a bias on the result of the overall combination
  is introduced}. The reason is that the prescription does not
  guaranty {\em statistical sufficiency}, whose importance is reminded
  in this script, written with a didactic spirit, with some historical notes
  and with a language to which most physicists are accustomed. 
  The conclusion contains general remarks on the effective
  presentation of the experimental findings and a pertinent {\em puzzle}
  is proposed in the Appendix.
\end{abstract}  
{\footnotesize
  \begin{flushright}
     {\sl ``Observations, for example, such as are distant from each other by
             an interval  of a few\\ days $[$\ldots$]$ are not
             to be used in the calculation
              as so many  different positions,
            but it would \\ be better to derive 
            from them a single place, which would be, as it were, 
             a mean among all,\\ admitting, therefore, much greater
             accuracy than single observations considered separately.'' } \\
{(F.C. Gauss, transl. by C.H. Davies)}\\
\end{flushright}
}

\section{Introduction}
As reminded by Gauss' opening quote, the reason why we
combine {\em ``single observations''}
is to have an equivalent one {\em ``of greater accuracy
  than the single observations''}. In fact, 
accidental errors tend to
cancel,\footnote{
    {\sl  ``It is of the greatest importance, that the several positions 
              of the heavenly body on which it is proposed to base the orbit,
              should not be 
             taken from single observations, but, if possible, 
             from several so combined that the accidental errors might,
             as far as may be, mutually destroy
             each other.''}\,\cite{Gauss_trasl}
           }
if the observations are
independent, and small systematic errors do too,
if the measurements are performed using different devices
and the data are analyzed by different
methods.
The simplest combination of the individual results
is the arithmetic mean. But\ldots -- and let speak    
Gauss again~\cite{Gauss_trasl} 
\begin{quote} {\sl 
    ``But if it seems that the same degree of accuracy cannot be attributed
    to the several observations, let us assume that the
    degree of accuracy in each may be considered proportional to the
    numbers $e$, $e'$,  $e''$, $e'''$, etc. respectively, that is,
    that errors reciprocally proportional to these numbers could have been
    made in the observations with equal facility; the, according to
    the principles to be propounded below, the most probable
    mean value will no longer be the simple arithmetic mean,
    but
    $$ \frac{e e \delta + e' e' \delta' +e'' e'' \delta'' +
      e''' e''' \delta''' + \mbox{etc} }
    {e e + e' e'  +e'' e'' +
      e''' e'''  + \mbox{etc}  }\,,\mbox{''}
    \hspace{1cm}\mbox{(G1)}
    $$
  }
\end{quote}
that is, in modern notation,
$ \left(\sum_ie_i^2\,\delta_i\right)/\left(\sum_ie_i^2\right)$,
in which we recognize the well known weighted average, provided $e_i=1/\sigma_i$.
Then, later on (the conversion to modern notation is now straightforward),
\begin{quote} {\sl 
    ``The degree of precision to be assigned to the mean found
    as above will be [\ldots]
     $$ \sqrt{}(e e + e' e'  +e'' e'' +
    e''' e'''  + \mbox{etc})\,;  \hspace{2.0cm}\mbox{(G2)}  $$
    so that four or nine equally exact observations are required,
    if the mean is to possess a double or a triple accuracy.''
   }
\end{quote}   
The advantage of having `distilled' the many observations
into a  single, equivalent one is that further calculations
are simplified, or made feasible at all,
especially in the absence of powerful computers.
But this idea works if there is no, or little,
loss of information.\footnote{For example Gauss discusses the implications
  of averaging observations over days or weeks, during which
  the heavenly body has certainly changed position in the elapsed time.
  But the mean position in the mean time can be considered as an
  equivalent point in space and time, and a few of them, far apart,
  would be enough to determine the orbit parameters.
}
 This leads us
to the important concept of {\em statistical sufficiency}, which will
be reminded in Sec.~\ref{sec:sufficiency} for the well
understood case of Gaussian errors.

There  is then the question of what to do in the case in which
the individual results
`appear' to be in mutual disagreement. The reason of the quote marks
has been discussed in Ref.~\cite{sceptical2019},
of which this work is a kind of appendix.
In short, they
are a reminder, if needed, of the fact that, rigorously speaking, we can never
be absolute sure that the `discrepancies' are not just due to
statistical fluctuations.
A way to implement our doubts has been shown
in Ref.~\cite{sceptical2019} and it consists in 
modifying
the probabilistic model relating {\em causes}
(the parameters of the model, first and foremost 
the `true' value of the quantity we aim to infer, although
with uncertainty)
and {\em effects} (the empirical observations),
adding some {\em extra-causes} (additional parameters of the model)
which might affect the observations. 
All quantities of interest are then embedded in a kind of network, a graphical
representation of which can be very helpful
to grasp its features.\footnote{The network
  of all quantities
  involved in the model is also known
  as {\em Bayesian network} for two reasons: {\em degrees of belief}
  are assigned to all variables in the game (even to the observed
  ones, meant as conditional probabilities depending on the
  value of the others); {\em inference} and
  {\em forecasting} (the names are related
  to the purposes of our analysis -- from the probabilistic point of
  view there is no difference) are then
  made by the use of probability rules, in particular
  the so called {\em Bayes' rule}, without the
  need to invent prescriptions or {\em ad hoc} `principles'.
}

Traditionally, at least in particle physics, a different approach is
followed. The degree of disagreement is quantified by
the $\chi^2$ of the differences between
the individual results and the weighted average.
Then, possibly, the standard deviation of the weighted mean is enlarged
by a factor $\sqrt{\chi^2/\nu}$. The rationale of the {\em prescription},
as a fast and dirty rule to get an rough idea of the range
of the possible values of the quantity of interest,
is easy to understand. 
However, there are several reasons of dissatisfaction, as discussed later on in
 Sec.~\ref{sec:chi2_su_nu}, 
and therefore
this simplistic scaling should be used with some care, and definitely avoided
in those cases in which the outcome
is critical for fundamental physics issues.\footnote{This
  is not the case, at the moment,
  for the charged kaon mass, as commented in  Ref.~\cite{sceptical2019},
  especially footnote 19.}
  Moreover, it will be shown how the outcome can be biased
  if the prescription is first applied to a sub-sample of the
  individual results and,
  subsequently, the partial result is combined
  with the remaining ones using the same
  rule. The conclusions will also contain
  some general considerations on how an experimental result should be presented
  in order to use it at best a) to confront it with theory;
  b) to combine it with other results concerning
  the same physics quantity; c) to `propagate' it
  into other quantities of interest.

  Finally, a puzzle is proposed in the Appendix in order to show that
  independent Gaussian errors are not a sufficient condition 
  for the standard weighted average.
  
\section{From a sample of individual observations to a
couple of numbers: the role of 
statistical sufficiency}\label{sec:sufficiency}

\begin{figure}[!t]
\begin{center}
\begin{tabular}{cc}    
  \hspace{0.5cm} \epsfig{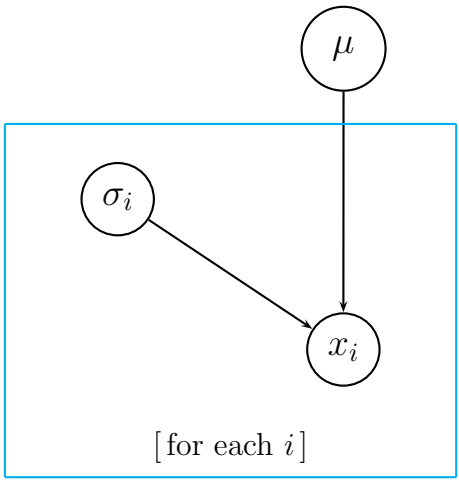}
\end{tabular}  
\end{center}  
\caption{\small \sf Graphical model behind the standard combination,
  assuming independent measurements of the same quantity, each
  characterized by a Gaussian error function with standard deviation
  $\sigma_i$}.
\label{fig:BN_ind_gaussianians}
\end{figure}

Let us restart from the Eq.~(5) of Ref.~\cite{sceptical2019},
based on the graphical model in Fig.~5 of the same paper,
reproduced here for the reader's convenience
as Eq.~(\ref{eq:fxy_1}) and Fig.~\ref{fig:BN_ind_gaussianians}.
\begin{eqnarray}  
  f(\underline{x},\mu,\,|\,\underline{\sigma})  & = &
  \left[ \prod_i f(x_i\,|\,\mu,\sigma_i)\right]\cdot f_0(\mu)\,
  \label{eq:fxy_1}
\end{eqnarray}
is the joint probability density function (pdf) of all the
quantities of interest,
with $\underline x = \left\{ x_1, x_2, \ldots \right\}$. The standard deviations
$\underline \sigma = \left\{\sigma_1, \sigma_2, \ldots\right\}$
are instead considered just {\em conditions} of the problem.
The pdf $f_0(\mu)$
models our prior beliefs about the `true' value
of the quantity of interest (see Ref.~\cite{sceptical2019}
for details, in particular footnote 9). The pdf of $\mu$, also
{\em conditioned
 on $\underline{x}$}, is then, in virtue of a well known theorem of
probability theory,
\begin{eqnarray}
  f(\mu\,|\,\underline{x},\underline{\sigma}) &=& 
  \frac{f(\underline{x},\mu,\,|\,\underline{\sigma})}
  {f(\underline{x}\,|\,\underline{\sigma})}\,.
  \label{eq:fy_1}
\end{eqnarray}
Noting that, given the model and the observed values
$\underline{x}$,
the denominator is just a number, although in general
not easy to calculate,
and making use of Eq.~(\ref{eq:fxy_1}), we get
\begin{eqnarray*}
  f(\mu\,|\,\underline{x},\underline{\sigma}) &\propto&
  \left[ \prod_i f(x_i\,|\,\mu,\sigma_i)\right]\cdot f_0(\mu) %
\end{eqnarray*}  
Speaking in terms of {\em likelihood}, and ignoring
multiplicative factors,\footnote{This is related to the 
  so called {\it Likelihood Principle}, which is consider
  a good feature by frequentists, 
though not all frequentistic methods 
respect it\,\cite{Zech0}. 
In practice, it says that the result of an inference should not
depend on multiplicative factors of the likelihood functions.
This `principle' arises automatically in a probabilistic
(`Bayesian') framework.}
we can rewrite the previous equation as
\begin{eqnarray*}
  f(\mu\,|\,\underline{x},\underline{\sigma}) &\propto&
   \left[ \prod_i {\cal L}(\mu\,;\,x_i,\sigma_i) \right]\cdot f_0(\mu)\,,
\end{eqnarray*}
that is, indeed, the particular case, valid for independent
observations $x_i$, of the more general form
\begin{eqnarray*}
  f(\mu\,|\,\underline{x},\underline{\sigma}) &\propto&
  {\cal L}(\mu\,;\,\underline{x},\underline{\sigma})  \cdot f_0(\mu)\,,
\end{eqnarray*}
since, under condition of independence,
$ {\cal L}(\mu\,;\,\underline{x},\underline{\sigma}) =  \prod_i {\cal L}(\mu\,;\,x_i,\sigma_i)\,.$
\begin{itemize} 
\item The inference depends on the product of likelihood and prior
  (note `;' instead of `$|$' in the notation,
  to remind that in `conventional statistics'
  ${\cal L}$ is simply a mathematical
  function of $\mu$, with parameters $\underline{x}$ and
  $\underline{\sigma}$); 
\item if the prior is `flat',\footnote{I refer again to footnote 9
  of Ref.~\cite{sceptical2019}, reminding that
  the {\em Princeps Mathematicorum}
  derived the `Gaussian' as the error function such that the
  maximum of the posterior for a flat `prior' (explicitly stated)
  had a maximum corresponding to the arithmetic average of the observations,
  provided they were independent and they had
  {\em ``the same level of accuracy''}.}
then the inference is determined by the likelihood,
\begin{eqnarray*}
  f(\mu\,|\,\underline{x},\underline{\sigma}) &\propto&
      {\cal L}(\mu\,;\,\underline{x},\underline{\sigma})\,.
\end{eqnarray*}
In particular,
the most probable value\footnote{Citing again
  Gauss, he wrote explicitly of most
  probable value of `$\mu$', under the hypothesis that
  before the experiment all its values were equally
  probable~\cite{Gauss_trasl}.}
(`mode') of $\mu$ is the value which maximizes
  the likelihood;\footnote{Hence the famous {\em Maximum likelihood}
    `principle', which is then nothing but a simple case of the more general
    probabilistic approach. (But the value
    that maximizes the likelihood is not necessarily the best
    value to report, as it will be commented in the conclusions.)
  }
\item in the case of independent Gaussian error functions
  the likelihood can be rewritten, besides multiplicative factors,
  as
  \begin{eqnarray}
    {\cal L}(\mu\,;\,\underline{x},\underline{\sigma}) =
    \prod_i {\cal L}(\mu\,;\,x_i,\sigma_i)
    &\propto& \prod_i \exp\left[-\frac{(x_i-\mu)^2}{2\,\sigma_i^2}\right]
    \nonumber \\
    &\propto&  \exp\left[-\sum_i\,\frac{(x_i-\mu)^2}{2\,\sigma_i^2}\right]
   \label{eq:sum_squares} \\
     &\propto&  \exp\left[-\chi^2/2\right]\,, \nonumber
  \end{eqnarray}
  having recognized  the sum in the exponent
  as $\chi^2 = \sum_i(x_i-\mu)^2/\sigma_i^2$:
  under the hypotheses and the approximations of this model the most probable
  value of $\mu$ can then also be obtained by
  minimizing $\chi^2$;\,\footnote{And  it is easy to recognize,
   in this sub-case of the general
    probabilistic
    approach, another famous `principle'.}
\item going through the steps from Eqs.~(7)-(12) of
  Ref.~\cite{sceptical2019} and, under
  the assumptions stated in the previous items,
  we can further rewrite the Eq.~(\ref{eq:sum_squares}) as
\begin{eqnarray}
    \exp\left[-\sum_i\,\frac{(x_i-\mu)^2}{2\,\sigma_i^2}\right] &\propto&
   \exp\left[- \frac { (\mu-\overline{x})^2}{2\,\sigma_C^2}\right] \,,
\end{eqnarray}
  where
\begin{eqnarray}
  \overline{x} &=& \frac{\sum_i\,x_i/\sigma_i^2} {\sum_i 1/\sigma_i^2}
  \label{eq:media_pesata}\\
  \frac{1}{\sigma_C^2} &=& \sum_i \frac{1}{\sigma_i^2}\,,
  \label{eq:sigma_media}  
\end{eqnarray}
in which we recognize Gauss' Eqs.~(G1) and (G2).
In terms of likelihoods,  
\begin{eqnarray}    
    {\cal L}(\mu\,;\,\underline{x},\underline{\sigma})& \propto&
    \prod_i {\cal L}(\mu\,;\,x_i,\sigma_i) \propto
    {\cal L}(\mu\,;\,\overline{x},\sigma_C)\,,
    \label{eq:sufficiency_media}
  \end{eqnarray}
\end{itemize}
Equation~(\ref{eq:sufficiency_media}) is an important result, related
to the concept of
{\em  statistical sufficiency}: the inference is exactly the same
if, instead of using the detailed information provided by
$\underline{x}$ and ${\underline{\sigma}}$, we just
use the weighted mean $\overline{x}$ and its
standard deviation $\sigma_C$, {\em as if $\overline x$ were
  a single equivalent observation of $\mu$ with a Gaussian error function
} with ``degree of accuracy''~\cite{Gauss_trasl} $1/\sigma_C$ --
this is exactly the result Gauss was aiming in Book 2, Section 3 of
Ref.~\cite{Gauss_trasl}, reminded in the opening quote
and in the introduction
of Ref.~\cite{sceptical2019}.

Moreover we can split the sum of Eq.(\ref{eq:sum_squares})
in two contributions,
from $i=1$ to $m$ (arbitrary) and from $m+1$ to $n$, thus having
 \begin{eqnarray}
   \exp\left[-\sum_i\,\frac{(x_i-\mu)^2}{2\,\sigma_i^2}\right]
   &=& \exp\left[-\sum_{i=1}^{m}\,\frac{(x_i-\mu)^2}{2\,\sigma_i^2} 
     -\sum_{i=m+1}^{n}\,\frac{(x_i-\mu)^2}{2\,\sigma_i^2}  \right]\,.
   \nonumber
  \end{eqnarray}
Going again through the steps from Eq.(7) to Eq.(12) of 
Ref.~\cite{sceptical2019} we get
\begin{eqnarray}
  \exp\left[-\sum_i\,\frac{(x_i-\mu)^2}{2\,\sigma_i^2}\right] &\propto&
   \exp\left[- \frac{ - 2\,\overline{x}_A\,\mu
       + \mu^2}{2\,\sigma_{C_A}^2} -
     \frac{ - 2\,\overline{x}_B\,\mu
       + \mu^2}{2\,\sigma_{C_B}^2}
     \right]  \nonumber
\end{eqnarray}  
where
\begin{eqnarray}
  \overline{x}_{A} &=& \frac{\sum_{i=1}^{m}\,x_i/\sigma_i^2}
  {\sum_{i=1}^{m} 1/\sigma_i^2}
  \label{eq:media_pesata_A}\\
  && \nonumber \\
  \sigma_{C_A}^2 &=& \frac{1}{\sum_{i=1}^{m} 1/\sigma_i^2}\,.
  \label{eq:sigma_media_A}\\
 && \nonumber \\  
  \overline{x}_B &=& \frac{\sum_{i=m+1}^{n}\,x_i/\sigma_i^2}
  {\sum_{i=m+1}^{n} 1/\sigma_i^2}
  \label{eq:media_pesata_B}\\
  && \nonumber \\
  \sigma_{C_B}^2 &=& \frac{1}{\sum_{i=m+1}^{n} 1/\sigma_i^2}\,.
  \label{eq:sigma_media_B}    
\end{eqnarray}
It follows, writing the right hand side as product of exponentials
and {\em complementing} each of them~\cite{sceptical2019}, 
\begin{eqnarray}
 \exp\left[- \frac { (\overline{x}-\mu)^2}
      {2\,\sigma_C^2}  \right]  &\propto&
    \exp\left[- \frac { (\overline{x}_{A}-\mu)^2}
      {2\,\sigma_{C_A}^2}  \right] \cdot
     \exp\left[- \frac { (\overline{x}_{B}-\mu)^2}
                                          {2\,\sigma_{C_B}^2}  \right] \\
   &\propto&
    \exp\left[- \frac { (\overline{x}_{A}-\mu)^2}
      {2\,\sigma_{C_A}^2} - \frac { (\overline{x}_{B}-\mu)^2}
                                          {2\,\sigma_{C_B}^2}  \right]\,,
\end{eqnarray}  
that is, in terms of likelihoods, 
\begin{eqnarray}
  {\cal L}(\mu\,;\,\overline{x},\sigma_C)  &\propto&
  {\cal L}(\mu\,;\,\overline{x}_A,\sigma_{C_A}) \cdot
   {\cal L}(\mu\,;\,\overline{x}_B,\sigma_{C_B})
  \\
   &\propto&
    {\cal L}(\mu\,;\,\overline{x}_A,\sigma_{C_A},
                   \overline{x}_B,\sigma_{C_B}) 
\end{eqnarray}
The result can be extended to averages of averages, that is
\begin{eqnarray}
    {\cal L}(\mu\,;\,\overline{x}_A,\sigma_{C_A},
    \overline{x}_B,\sigma_{C_B})
    &\propto&
    {\cal L}\left(\mu\,;\,{\overline{x}}^{(G)},{\sigma_{C}}^{(G)}\right)\,,           
\end{eqnarray}
where 
\begin{eqnarray}
  \overline{x}^{(G)} &=& \frac{\overline{x}_A/\sigma_{C_A}^2
    + \overline{x}_B/\sigma_{C_B}^2}
           {1/\sigma_{C_A}^2 + 1/\sigma_{C_B}^2} \\
  && \nonumber \\
           \frac{1}{{\sigma_{C}^2}^{(G)}} &=& \frac{1}{\sigma_{C_A}^2} +
           \frac{1}{\sigma_{C_B}^2}\,.
  \label{eq:sigma_media_G}
\end{eqnarray}
%
The property can be extended further to many partial averages,
showing that the {\em inference 
 does not depend on whether we use the individual
 observations, their weighted average or even the grouped weighted
 averages},  or  the weighted average of the grouped averages.
 This is one of the `amazing' properties
 of the Gaussian distribution, which simplifies our work
 when it is possible to use it.
  But there no
  guarantee that it works in general, and it should be
  then proved case by case.

 \section{Possibly discrepant results}\label{sec:discrepant_results}
 Well known features of the combination 
 of measurements by the weighted
 average, reminded in the last section, are that i) the combined result
 has a ``degree of precision''~\cite{Gauss_trasl}
 higher than each of the individual contributions
 or, in terms of standard deviations, $\sigma_C < \sigma_i$; 
 ii) the resulting standard deviation
 does not depend on the spread of the individual
 results around the mean value;
 iii) the error model of the `equivalent observation'
 remains Gaussian. However, it is a matter of fact that,
 although from the probabilistic point of view
 there is no contradiction with the basic assumptions, since
 patterns of individual results `oddly' scattered around their average
 have some chance to occur, we  sometimes {\em suspect} that
 there it might be something `odd' going on. That is, we
 tend to doubt on the validity of the simple model of Gaussian errors with
 the declared ``degrees of precision''. 
 (But someone might start to worry 
 too early,\footnote{It is rather
   well known that the human mind has problems when 
   dealing with randomness. For example, if you ask a person
   to write, at random, a long list of 0's and 1's,
   she will tend to `regularize' the series, which will then
   contain only short sequences of 0's and 1's,
   contrary to what happens rolling a coin, or using a
   (pseudo-)random generator. As an interesting book that
   discusses, among others, this {\em experimental fact},
   Ref.~\cite{Kahneman} is recommended.
 } sometimes even driven by
 {\em wishful thinking}, that is she {\em hopes},
 rather than believes, that the reason of disagreement might be
 caused by new phenomenology or violation
 of fundamental laws of physics
 \cite{ProbablyDiscovery,WavesSigmas}.) 

  \begin{figure}[!t]
\centering\epsfig{file=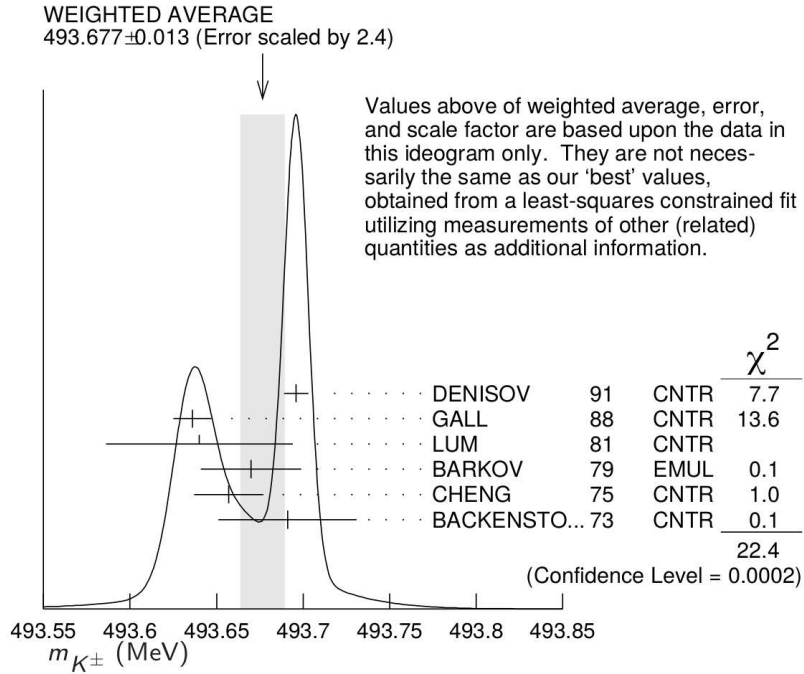,clip=,width=0.7\linewidth}
\caption{\small \sf Charged kaon mass from several experiments
  as summarized by the PDG~\cite{PDG2019}. Note that besides
  the `error' of 0.013\,MeV,
  obtained by a $\times 2.4$ scaling,
  also an `error' of  0.016\,MeV is provided,
  obtained by a $\times 2.8$ scaling. The two results are
  called `OUR AVERAGE' and 'OUR FIT', respectively~\cite{PDG2019}.
}
\label{fig:PDF2019}
\end{figure}   
 In the case we have serious suspicions about the
 presence of other effects, then
 we should change our model, make a new analysis and accept its outcome
 in the light of clearly stated hypotheses and
 conditions~\cite{sceptical2019,sceptical1999,Dose}.
 As a result, not only the overall `error'\footnote{The
   quote marks are to remind
    that they refer, more precisely,
    to {\em standard uncertainty},
    referring the nouns {\em error} and {\em uncertainty}
    to different concepts\,\cite{ISO,ISOD}.}
 should change, but also the 
 shape of the final distribution should, since there is no strong
 reason to remain Gaussian.
 For example, the final distribution
 might be skewed or even multimodal~\cite{sceptical2019}, 
 as it should be desirable
 if the pattern of individual measurements
 suggest so. In particular,
 the most probable value ({\em mode}) will differ from the
 average of the distribution and from the {\em median}.
 Instead, traditionally, only the
 `error' is enlarged by an {\em arbitrary} factor
 depending on the frequentistic  `test variable' $\chi^2$,
 namely $\sqrt{\chi^2/\nu}$, where $\nu$ stands for the number of
 {\em degrees of freedom}.  
 But the central value is
 kept  unchanged and the interpretation of the result, explicitly stated
 or implicitly assumed so
 in subsequent analyses by other scientists,
 remains Gaussian.\footnote{For example, the bimodal
   curve shown in the ideogram of Fig.~\ref{fig:PDF2019}
   is a curious linear combination of Gaussians
   \underline{of}
   $m_{K^\pm}$, although {\em from a frequentistic point
     of view one should not be allowed to attribute probabilities, and hence
     pdf's, to true values}. And there are even frequentistic `gurus'
   who use probability in quote marks, without explaining the reason but
   because they are aware that could not talk about probability, when they write
   ``{\em When the result of a measurement of a physical quantity
     is published as $R=R_0\pm\sigma_0$ without further explanation, it is implied
     that R is a gaussian-distributed measurement with mean $R_0$
     and variance $\sigma_0^2$. This allows one to calculate
     various confidence intervals of given ``probability'', i.e.,
   the ``probability'' $P$ that the true value of $R$ is withing a given interval.}''\,\cite{James-Roos}.
 } 

\begin{table}[!t]
  \begin{center}
{\footnotesize
  \begin{tabular}{|c|l|c|c|c|c|}
    \hline
 &   \multicolumn{1}{|c|}{Authors} & pub. year & central value $[d_i]$ & uncertainty $[s_i]$\\
 $i$            &        &   &    (MeV)      &    (MeV)    \\
    \hline
 $1$ &  G. Backenstoss et al. \cite{Kmass73}   & 1973 &  493.691 & 0.040 \\
 $2$ &     S.C. Cheng et al. \cite{Kmass75}        & 1975 & 493.657  & 0.020 \\
 $3$ &     L.M. Barkov et al.\cite{Kmass79}       & 1979 &  493.670  &  0.029 \\
 $4$ &     G.K. Lum et al. \cite{Kmass81}       & 1981   &  493.640  &   0.054 \\
 $5$ &     K.P. Gall et al. \cite{Kmass88}      &  1988  & 493.636   &    0.011 (*) \\
 $6$ &     A.S. Denisov et al. \cite{Kmass91}   &  1991  & 493.696  &  $[${\em 0.0059}$]$ \\
   &  \& {\em Yu.M. Ivanov} \cite{Kmass92}  & 1992  & $[${\em same}$]$&  0.007 (**)\\
  \hline
\end{tabular}
\caption{\footnotesize \sf Experimental values of the charged kaon mass
  used in the numerical example, limited to those taken into account
  by the 2019 issue of PDG~\cite{PDG2019}. $[$(*) `error' already scaled
    by a factor $\times 1.52$ due to the $\sqrt{\chi^2/\nu}$ prescription
    (see text).\ (**) Value accepted by the PDG.$]$
}
}
\end{center}
\label{tab:masseK_PDG}
\end{table}
 As a practical example, let us take the results concerning the
 charged kaon mass of Fig.~\ref{fig:PDF2019} and Tab.~1, 
 as selected by the PDG\,\cite{PDG2019}.
 \begin{figure}[!t]
   \centering\epsfig{file=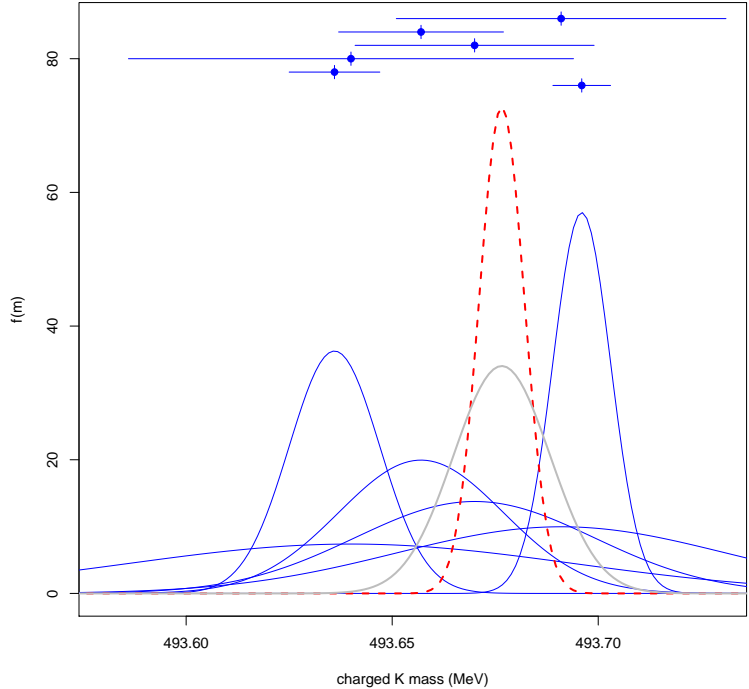,clip=,width=0.65\linewidth}
\caption{\small \sf Graphical representation of the results on the charged kaon mass
  of Tab.~1 
  (solid blue Gaussians).
  The dashed  red  Gaussian shows the result
  of the standard combination obtained by the weighted average.
  The solid gray Gaussian, centered with the dashed red one,
  shows the broadening due to  
  the $\sqrt{\chi^2/\nu}$
  prescription (see text).}
\label{fig:NaiveCombination} 
 \end{figure}
From  the weighted average and its
standard deviation
we get $493.6766 \pm 0.0055\,$MeV,
shown in Fig.~\ref{fig:NaiveCombination} by the dashed red Gaussian
 (the solid blue Gaussians depict the results of the six
 results of Tab.~1. 
 Comparing the individual results with the weighted average
  we calculate
a $\chi^2$ of 22.9, and hence a scaling factors of 2.14, getting
then
$493.677 \pm 0.012\,$MeV, reported
on the same figure by the solid gray Gaussian below the 
dashed one.\footnote{Besides rounding, the numbers slightly differ from those
  of Ref.~\cite{PDG2019}, having applied there some
  {\em ad hoc} selections. But this does not
  change the essence of the message this note desires to convey.}
The two results are reported also in the entry A of
the summary table 3.

\section{Reasons of dissatisfaction with the
  $\chi^2$-motivated scaling prescription}\label{sec:chi2_su_nu}
As we see in  Fig.~\ref{fig:NaiveCombination}, the Gaussian
widened by the $\sqrt{\chi^2/\nu}$ prescription does not capture
the picture offered by the ensemble of the individual results.
In fact, the mass values
preferred by the combined result are still distributed
symmetrically around the weighted
average.\footnote{Then the PDG adds a more intriguing curve
  n the reported `ideograms'\cite{PDG2019} (see e.g.
  Fig.~1 of Ref.~\cite{sceptical2019}). But this non-Gaussian
  curve 
  has no probabilistic meaning,
  as discussed in \cite{sceptical2019}, and, anyway,
  it is not used to draw no quantitative
results, as far as I understand (and I hope\ldots).}  

More in general, 
the scaling factor
is at least suspect.
This is because it is well known that the
$\chi^2$ distribution does not scale with $\nu$ and therefore,
while a $\chi/\nu = 2$, for example, is quite in the norm
for $\nu$ equal to 2, 3  or 4 (even a strict frequentist would admit
that the resulting p-values of 0.14, 0.11 and 0.09, respectively,
are nothing to worry), things get different for $\nu$ equal to 10,
20 or 30 (p-values of 0.029, 0.005 and 0.0009, respectively).
Moreover, I am not aware of cases in which the standard
deviation of  the weighted average was scaled down,
in the case that $\sqrt{\chi^2/\nu}$ was smaller
than one.\footnote{It seems that people
  are worried only if the `errors' appear small.
  But, as also stated by the ISO Guide~\cite{ISO}, 
  {\sl ``Uncertainties in measurements''} should
  be {\sl ``realistic rather than safe''}.
  In particular the method recommended by the ISO Guide
  {\sl ``stands [\ldots] in contrast to certain older methods that have
    the following two ideas in common:
    \begin{itemize}
    \item The first idea is that the uncertainty reported should be
      `safe' or `conservative [\ldots] In fact, because the evaluation of the
      uncertainty of a measurement result is problematic, it was often
      made deliberately large. [\ldots]''
    \end{itemize}    
  }
} 

But there is another subtle issue with the method,
which I have realized only very recently, going through the details
of the charged kaon mass measurements:  if
the prescription is applied to a sub-sample of results
and then to all them
(taking for the sub-sample weighted average and scaled standard deviation),
then {\em a bias is introduced in the final result} with respect to when
all results were taken individually.
This is because the summary provided by such a prescription
{\em is not a sufficient statistics}.

The lowest, high precision  mass value of $493.636 \pm 0.011$
(see Tab.~1 and Fig.~\ref{fig:NaiveCombination})
come in fact from the combination, done directly by the experimental
team~\cite{Kmass88} applying the $\sqrt{\chi^2/\nu}$ prescription.
Without this scaling, the four individual results,
reported in Tab.~2, 
\begin{table}[!t]
\begin{center}
{\footnotesize
  \begin{tabular}{|c|l|c|c|c|c|}
    \hline
 &   \multicolumn{1}{|c|}{Authors} & pub. year &  $[d_i]$ & $[s_i]$\\
 $i$            &        &   &    (MeV)      &    (MeV)    \\
    \hline
$5_a$ &     K.P. Gall et al. \cite{Kmass88}      &  1988  & 493.675 &  0.026   \\
$5_b$ &                                          &        & 493.631 &  0.007   \\
$5_c$ &                                          &        & 493.806 &  0.095   \\
    $5_d$ &                                          &        & 493.709 &  0.073   \\
    \hline
$5$ &     K.P. Gall et al. \cite{Kmass88}      &  1988  & 493.636   &    0.011 \\    
  \hline
  \end{tabular}
  }
\caption{\footnotesize \sf Individual results reported by \cite{Kmass88},
  together with their combination.
}
\end{center}
\label{tab:masseK_Gall}
\end{table}
 \begin{table}
  {\footnotesize
   \begin{center}
   \begin{tabular}{c|c|c}
   & data set & $m_{K^\pm}$/MeV \\  
     \hline
 \multirow{2}{*}{A)} &     $\{1,2,3,4,5,6\}$   & $493.6766\pm0.0055$ \\
    & $\{1,2,3,4,5,6\}_S$   & $\mathbf{493.677}\pm0.012$  \\
\hline
 \multirow{2}{*}{B)} & $\{5_a, 5_b, 5_c, 5_d\}$ & $493.6355\pm 0.0067$ \\
     & $\{5_a, 5_b, 5_c, 5_d\}_S$ & $493.636\pm 0.010$ \\
 \hline
 \multirow{2}{*}{C)} &   $\{1,2,3,4,5_a,5_b,5_c,5_d,6\}$ & $493.6644\pm 0.0046$ \\
     & $\{1,2,3,4,5_a,5_b,5_c,5_d,6\}_S$ & $\mathbf{493.664}\pm 0.011$ \\
     \hline
 \multirow{2}{*}{D)} &   $\{1,2,3,4,5_a,5_b,5_c,5_d\}$   & $493.6404\pm 0.0061$  \\
     & $\{1,2,3,4,5_a,5_b,5_c,5_d\}_S$   & $493.6404\pm 0.0076$ \\
     \hline 
 \multirow{2}{*}{E)} &   $\{\{1,2,3,4,5_a,5_b,5_c,5_d\}_S,6\} $   & $493.6705\pm 0.0051$  \\
      & $\{\{1,2,3,4,5_a,5_b,5_c,5_d\}_S,6\}_S $   & $493.671\pm 0.028$  
   \end{tabular}
   \caption{\small {\sl Combinations of the individual results
       of Tabs. 1   and 2. The
       subscript $S$ means that the $\sqrt{\chi^2/\nu}$ scaling prescription
       has
       been applied to standard deviation of the weighted average.
       In particular,  note that $\{1,2,3,4,5,6\}_S$ is the same as
       $\{1,2,3,4,\{5_a,5_b,5_c,5_d\}_S,6\}_S$.
     }}
 \end{center}
}
\label{tab:confronto}
\end{table}
had given a weighted average of $493.6355 \pm 0.0067\,$MeV,
with a $\chi^2$ of 7.0.
Now it is true that
  $\chi^2/\nu$ is equal to 2.32, but this is not a reason to worry,
  being $\nu=3$. In fact the {\em p-value}, calculated
  as $P(\chi^2\,|\,\nu=3) > 7.0$, 
  is 0.073, that is
  even above the (in-)famous 0.05 threshold~\cite{WavesSigmas}.

\begin{figure}[!t]
    \begin{center}
      \epsfig{file=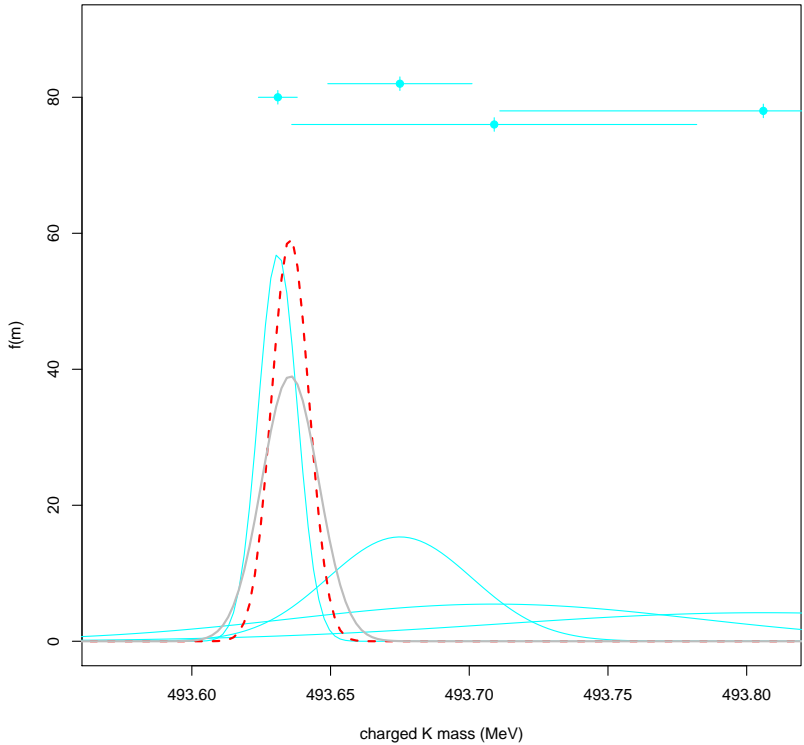,clip=,width=0.7\linewidth}
  \end{center}
    \caption{\small \sf
      Individual results of Ref.~\cite{Kmass88} (cyan solid lines),
      with the
      weighted  average with and without
      $\sqrt{\chi^2/\nu}$ scaling factor
      (same graphic notation of Fig.~\ref{fig:NaiveCombination}).
    }
  \label{fig:GallDetails}
  \end{figure}

  Nevertheless,
  if we apply to the standard deviation a scaling factor of
  $\sqrt{2.32} = 1.52$, then
  we get  $493.636 \pm 0.010\,$MeV (the difference between this value
  of 0.010\,MeV
  and 0.011\,MeV of Tabs.~1 and 2
  could be just due to rounding of the individual values).
  The result is shown in 
  Fig.~\ref{fig:GallDetails}, together with the individual results
  that enter the analysis (see also entry B of the summary table 3).
  
  It is interesting to see what we get if we use
  the nine individual points, i.e.
  1, 2, 3, 4 and 6 of Tab.~1, together
  with $5_a$,  $5_b$, $5_c$ and   $5_d$ of Tab.~2.
 \begin{figure}[!t]
    \begin{center}
      \epsfig{file=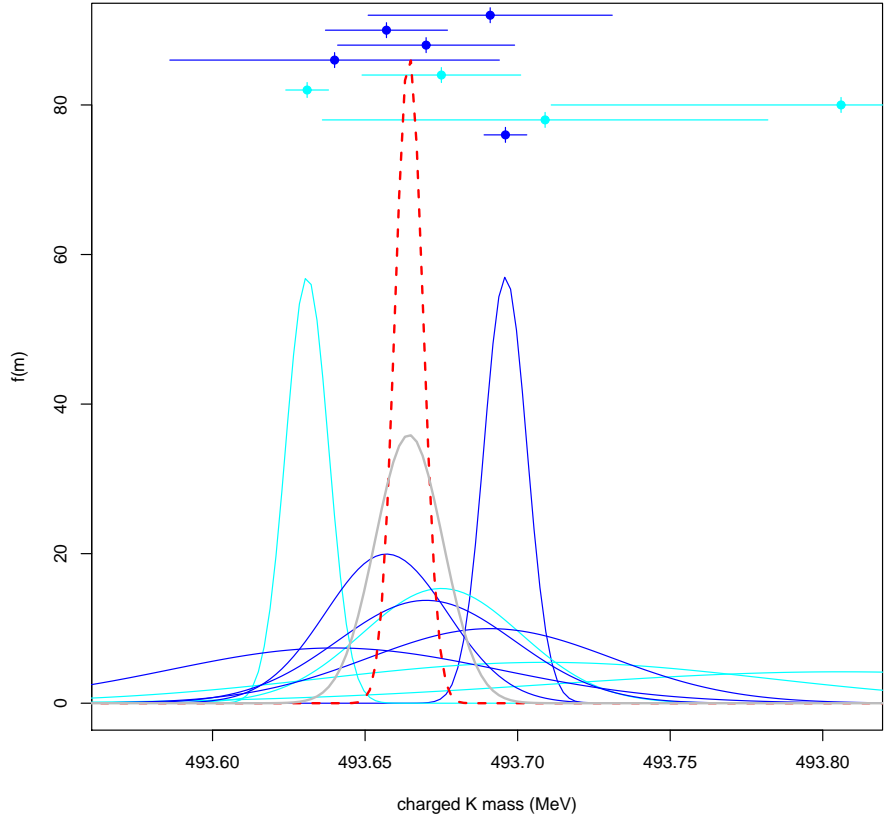,clip=,width=0.77\linewidth}
%
  \end{center}
    \caption{\small \sf
      Combination of the individual results of  Ref.~\cite{Kmass88}, together
      with the {\em other} results (i.e. excluding nr. $5$)
      of Tab.~1. For details of the graphic notation see
      the previous figures.
    }
  \label{fig:GlobalGallDetails}
  \end{figure} 
  The combined weighted average, shown in Fig.~\ref{fig:GlobalGallDetails}, 
 comes out right in the middle of the two most
 precise results, with little overlap with them. The average is 
 $493.6644\,$MeV, with standard deviation $0.0046\,$MeV, which becomes
 $0.011\,$MeV after the $\chi^2$ motivated scaling\footnote{The shift of the
   average almost in the middle of the two most precise results makes
   the contributions to the $\chi^2$ huge. Here are the differences
   of the individual measurements in unit of their standard deviations:
   $0.67,\, -0.37,\,0.19,\, -0.45,\, 0.41,\, -4.77,\, 1.49,\, 0.61,\, 4.52$,
   resulting in a $\chi^2$ of 46.7 and a consequent p-value of $1.7\times 10^{-7}$.
   Therefore, also to whom who are critical against
   p-values~\cite{ProbablyDiscovery,WavesSigmas},
   {\em in a case of this kind}, an alarm bell should sound~\cite{MSMM2019}.
   But to all persons of good sense a similar alarm,
   concerning the {\em frequentist solution} of the problem,
   should sound too (see Ref.~\cite{sceptical2019}
   for {\em a} sceptical alternative.)
 }
 of $\times 2.42$.
 As we can see, the central value differs by $-12\,$keV
 with respect from the one obtained above $[$\,see
   also section.~\ref{sec:discrepant_results} and entry C
   of the summary table 3\,$]$: the use of
 the pre-combined result of Ref.~\cite{Kmass88}
 {\em produces a bias of} $+12\,$keV {\em in the final result},
 that is
 comparable with the quoted `error'. 
 The reason is due to the fact
 that the  $\sqrt{\chi^2/\nu}$ {\em prescription
 used to enlarge the standard deviation does not hold sufficiency}.
 As a consequence, {\em the relevance of the ensemble
 of results of Ref.~\cite{Kmass88} gets reduced}.

\begin{figure} 
    \begin{center}
      \begin{tabular}{c}
        \epsfig{file=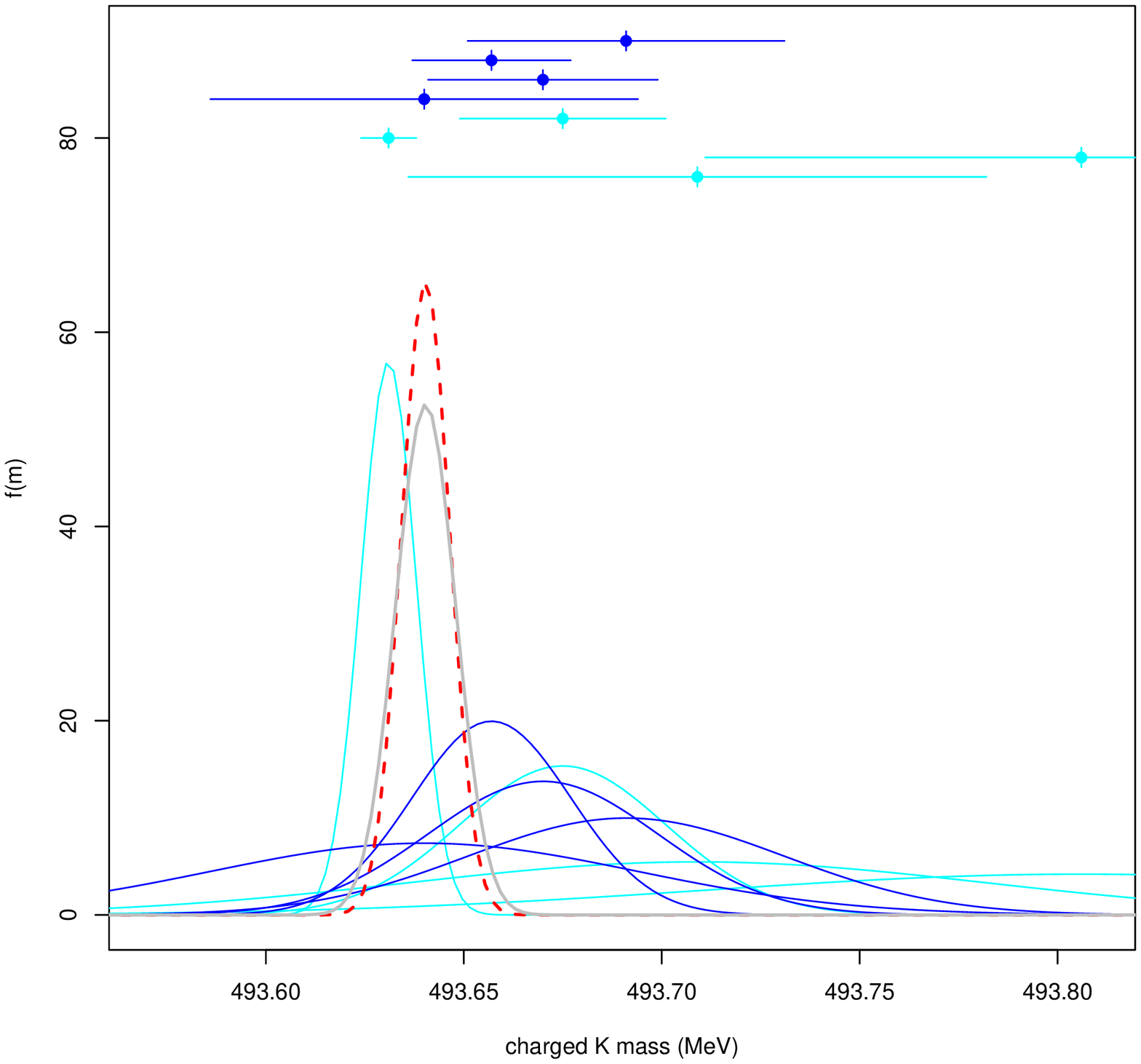,clip=,width=0.63\linewidth} \\
%
%
        \epsfig{file=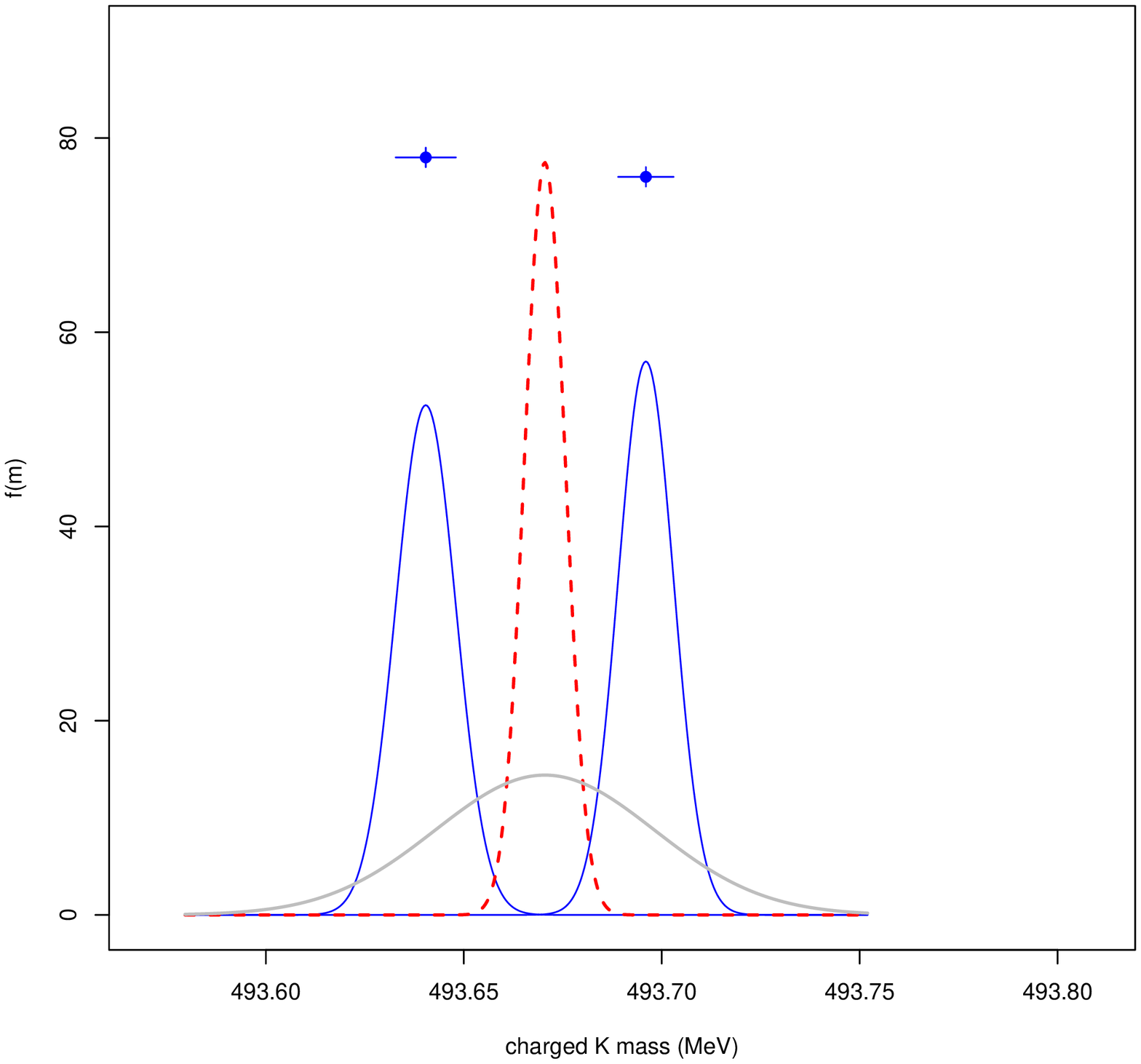,clip=,width=0.63\linewidth}
%
%
        \end{tabular}
  \end{center}
    \caption{\small \sf
      Example of arbitrary grouping of the results in before
      and after year 1990 (see text).
    }
  \label{fig:BeforeAfter1990}
  \end{figure}

 As a further example to show this effect on the same data,
 let us make the {\em academic exercise} of grouping
 the data in a different way. For example we first combine all results
 published before year 1990 (1-4,$5_a$-$5_d$, with references to
 Tabs.~ 1 and 2, 
 and include the most recent one (6 of Tab.~1) 
    in a second step.
    The outcome of the exercise is reported in Fig.~\ref{fig:BeforeAfter1990}
    and in the entries D and E of the summary table 3. 
  The weighted average of the eight results before year 1990
  (upper plot of Fig.~\ref{fig:BeforeAfter1990}
  and entry D in  Tab.~3) 
  gives $m_{K^\pm} = 493.6404\pm0.0061\,$MeV (dashed red line).
  The $\chi^2$ is equal
  to 10.8, producing a scaling factor of 1.24 and thus a modified
  result of $m_{K^\pm} = 493.6404\pm0.0076\,$MeV (solid brown line
   of Fig.~\ref{fig:BeforeAfter1990}
  and entry D in Tab.~3). 
  
  Combining this outcome with the 1991
  result~\cite{Kmass91,Kmass92}
  we get (lower plot of Fig.~\ref{fig:BeforeAfter1990}
  and entry E in Tab.~3) 
  a weighted average of 
  $m_{K^\pm} = 493.6705\pm 0.0051\,$MeV, but 
  with the very large $\chi^2$ of 29
  (p-value $0.74\times 10^{-7}$), thus yielding
  a $\times 5.4$ scaling factor and then a widened
  standard deviation of $28\,$keV. At least, contrary
  to the previous cases,
  this time the scaled standard deviation is able
  to cover both individual results, although
  an experienced physicist would suspect that
  {\em most likely} only one of the two  is
  correct. (In situations of this kind a `sceptical analysis'
    would result in a bimodal distribution, as shown in Fig.~4 of
    Ref.~\cite{sceptical1999}.)
  

\section{Conclusions}
It is self-evident that the
prescription of enlarging the standard deviation
of a weighted average by a $\chi^2$ motivated scaling factor
is often not able to capture the pattern of individual results,
in the case they show a sizable skewed distribution or
they tend to cluster in two different regions.
Instead, I was not aware of the fact that
the $\sqrt{\chi^2/\nu}$ scaling
might produce a bias, with respect to the average of
all individual results, if the scaling is first
applied to a sub-sample of results, and then
to all of them~\cite{sceptical2019}. This is due to the fact that the procedure
does not hold statistical sufficiency
and, therefore,
individual results should be used  
without pre-grouping.

However, this does not imply that the latter is the
correct way to proceed in the case the pattern
of the individual results is at odds with the
weighted average applied to all points. A more pondered
analysis should rather be performed in order to model our doubts,
as done e.g. in Ref.~\cite{sceptical2019}.
(In the case of the charged kaon mass, there is however
a curious compensation, such that the biased result comes out
to agree, at least in terms of central value and `error',
with that of the `sceptical analysis'~\cite{sceptical2019}.)

I would like to conclude with some remarks
concerning how to report an experimental result,
in perspective of its further uses. In fact,
a result is not an end in itself, as {\em Physics
and all Sciences are not just collections of facts}
(and even an experimental result is not a mere `fact',
since it is derived from many
empirical observations through {\em models relying on a web of
beliefs}\footnote{As the historian of science Peter Galison puts it,
{\sl ``Experiments begin and end in a matrix of beliefs. 
\ldots beliefs in instrument type, in programs of experiment
inquiry, in the trained, individual judgments about every local behavior
of pieces of apparatus.''~\cite{Galison}.}
\label{fn:Galison}}).

Focusing on pure science,
results are finally confronted with theoretical evaluations
(not strictly `predictions') in order to
classify in degree of belief the possible models 
describing how `the World works' 
(note that the acclaimed Popperian falsification
is an idealistic scheme that however
seldom applies in practice\,\cite{Vulcano,Wilczek}).
But in order to achieve
the best selective power, individual results are combined
together, as we have seen in this note.
Moreover a result could be {\em propagated} into other
evaluations, as it is, itself,
practically always based on other results, since it depends
on quantities which enter the theoretical model(s)
on which it relies (`principles of measurement'~\cite{ISO}),
including those who govern the ``pieces of apparatus'',
as reminded in footnote \ref{fn:Galison}. 

Therefore, it is important to provide, as outcome of an experimental
investigation, something that can be used at
best, even after years, for {\em comparison}, {\em combination}
and {\em propagation}. Fortunately there is 
something on which there is universal
consensus: the most complete information 
resulting from the
the empirical findings, concerning a quantity that can assume values
with continuity, is the so called
{\em likelihood function}.\footnote{In the case
there are only some possibilities, or we are interested
on how the data support an {\em hypothesis} over others,
the quantity to report is (are) the {\em likelihood ratio}(s),
as recently recognized also by the European Network of Forensic Science
Institutes (ENFSI)\,\cite{ENFSI}. Note that a likelihood function,
or likelihood ratios (also known as {\em Bayes factors}), cannot
be considered as `objective numbers', because they depend 
on the judgments of experts. This is also recognized by the
ENFSI Guidelines\,\cite{ENFSI}, which appears then rather
advanced with respect to naive ideals of objectivism that often
are  instead just a defense of established procedures and methods
supported by inertia and authority.
}
In fact, in the case of independent experiments
 reporting evidence
on the same physics quantity the {\em rule of
the combination} is straightforwards, as it results from probability
theory, without the need of {\em ad hoc} prescriptions:
just {\em multiply the individual likelihoods}.
It follows then that the likelihood (or its negative log)
should be described at best in a publication, as for example done
in Ref.\cite{ZeusContact}, in which several
negative log-likelihoods were
shown in figures and parameterized around their minimum
by suitable polynomials. 

Reducing the detailed information provided by the likelihood
in a couple of numbers does not provide, in general, an  effective and 
unbiased way to report the  result of the findings,
unless the likelihood is, with some degree of approximation, Gaussian.
Instead, if the likelihood is not Gaussian [or
  the $\chi^2$ is not parabolic, in those
cases in which the likelihood can be rewritten as $\exp(-\chi^2/2)$], then
reporting the value that maximize it, with an `error' related
to the curvature of its negative log at the minimum, or `asymmetric errors'
derived from a prescription that is only justified for a
Gaussian likelihood, is also an inappropriate way
of reporting the information contained in the findings.
This is because, when a result is given
in terms of $\hat{x}^{+\Delta_+}_{-\Delta_-}$, then 
$\hat x$ is  often used in further calculations, and the
$(\Delta_+,\Delta_-)$'s are `propagated' into further uncertainties
in `creative' ways, forgetting that the well know formulae
for propagations in linear combinations
(or in linearized forms) rely on probabilistic properties
of means and variances (and the Central Limit Theorem
makes the result Gaussian if `several' contributions are considered).
There are, instead, no similar theorems that apply to the
`best values' obtaining minimizing the $-\ln{\cal L}$ or the $\chi^2$
and to the (possibly asymmetric) `errors' obtained by the
$\Delta(-\ln{\cal L})$ and the $\Delta\chi^2$ rules, still
commonly used to evaluate `errors'.
Therefore these rules might produce biased results, directly
and/or in propagations~\cite{Asymmetric}.

Reporting the likelihood is also very important in the case
of `negative searches', in which a lower/upper
bound is usually reported. In fact, although there is no
way to combine the bounds (and so people often rely on the most stringent
one, which could be just due a larger fluctuation of the background
with respect to its expectation)
there are little doubts on how to
`merge' the individual (independent) likelihoods in a single
combined likelihood,
from which {\em conventional bounds} can be evaluated (see
Ref.~\cite{CLW1999} and chapter 13 of Ref.~\cite{BR}).

Finally, a puzzle is proposed in the Appendix, as a warning
on the use 
of the weighted average to combine results,
even if they are believed to be
independent and affected by Gaussian errors.

\mbox{} \\
It is indebted to Enrico Franco for extensive
discussions on the subject and comments on the manuscript.

\newpage
\section*{Appendix -- Independent Gaussian
  errors are not a sufficient condition to rely on the standard
  weighted average:  an instructive puzzle}
Imagine we have a sample of  $n$ observations, characterized
by independent Gaussian errors with unkown $\mu$,
associated to the true value of interest, and also unkown $\sigma$.
Our main interest is to infer $\mu$, but in this case also $\sigma$
needs to be estimated from the same sample. The `estimators' (to use
frequentistic vocabulary, to which most readers are most likely
familiar)
of  $\mu$ and  $\sigma$ are the aritmetic mean $\overline{x}$ and 
the standard deviation $s$ calculated from the sample,
respectively.\footnote{See e.g. Sec. 39.2.1 of Ref
  \cite{PDG2019},\\
  \url{http://pdg.lbl.gov/2019/reviews/rpp2019-rev-statistics.pdf}}
In particular the `error' on $\mu$ is calculated as
$s/\sqrt{n}$ (hereafter we focus on the determination
of $\mu$, although a similar reasoning and a related puzzle
concerns the determination of $\sigma$).

Now the question is what happens if we devide the samples in sub-samples,
`determine' $\mu$ from each sub-sample and then
combine the partial results. In order to avoid abstract
speculations, let us concentrate on the following simulated
sample:\footnote{The sample has been obtained
  with the Gaussian random number generator {\tt rnorm()}
  of the {\tt R} language\cite{R}, with the following commands\\
  {\tt
\mbox{}\ \ \ set.seed(20200102)\\
\mbox{}\ \ \ n = 16; mu = 3; sigma = 1\\
\mbox{}\ \ \ x  = rnorm(n, mu, sigma)}\\
so that {\em we} know the `true $\mu$'.
  (The random seed, used to make the
  numbers reproducible, was set to the date in which
  the sample was generated.) Mean and standard deviation
  are then calculated using {\tt R} functions:\\
  {\tt
    \mbox{}\ \ \ m = mean(x)\\
    \mbox{}\ \ \ s = sd(x)
    }
}
{\small \\ \mbox{} \\
\begin{tabular}{cccccccc}
  2.691952 & 2.805799 & 3.826049 & 1.908438, 
  3.844093 & 2.406228 & 5.176920 & 1.925284 \\
  1.688440 & 2.309165 & 3.046256 & 3.211285,
  2.302760 & 2.966700 & 2.301784 & 2.232128
\end{tabular}  
\\ \mbox{} \\ }
From the aritmetic average (2.7902) and the `empirical'
standard deviation (0.8970), we get
$$\mu^{(All)} = 2.790 \pm 0.224$$
(the exaggerate number of decimal digits, with respect
to reasonable standards, is only to make comparisons easier). 

Let us now split the values into two sub-samples (first and second
row, respectively). The `determinations' of $\mu$ are now
\begin{eqnarray*} 
  \mu^{(A)} &=& 3.073 \pm 0.399 \\
  \mu^{(B)} &=& 2.507 \pm  0.183\,.
\end{eqnarray*}
Combining then the two results calculating the weighted average
and its standard deviation, we get
\begin{eqnarray*}
  \mu^{(A\&B)} &=& 2.605 \pm  0.1660\,,
\end{eqnarray*}
sensibly different from $\mu^{(All)}$ calculated above. 

We can then split again the two samples (first four values and
second four values of each row), thus getting
\begin{eqnarray*} 
  \mu^{(A_1)} &=&  2.808 \pm 0.394 \\
  \mu^{(A_2)} &=&  3.338 \pm 0.736 \\
  \mu^{(B_1)} &=&  2.564 \pm 0.352 \\
  \mu^{(B_2)} &=&  2.451 \pm 0.173\,. 
\end{eqnarray*}
Combining the four partial results we get then
\begin{eqnarray*}
  \mu^{(A_1\&A_2\&B_1\&B2)} &=&  2.548  \pm 0.142 \,,
\end{eqnarray*}
different from $\mu^{(All)}$ and from $\mu^{(A\&B)}$.
\mbox{} \\\mbox{} \\
{\bf What is going on?} Or, more precisely,
what should be the combination rule such that
$\mu^{(All)}$, $\mu^{(A\&B)}$ and  $\mu^{(A_1\&A_2\&B_1\&B2)}$
would be the same?
(Sufficient hints are in the paper
and a note could possibly follow with a detailed treatement of the case.)
\end{document}